\documentclass[10pt]{IEEEtran}

\makeatletter
\def\ps@headings{%
\def\@oddhead{\mbox{}\scriptsize\rightmark \hfil \thepage}%
\def\@evenhead{\scriptsize\thepage \hfil \leftmark\mbox{}}%
\def\@oddfoot{}%
\def\@evenfoot{}}

\makeatother

\usepackage{amssymb,amsmath}
\usepackage{amsfonts}
\usepackage{array}
\usepackage{algorithmicx}
\usepackage[ruled,commentsnumbered]{algorithm2e}
\usepackage{subfigure}
\usepackage{mathrsfs}
\usepackage[T1]{fontenc}
\usepackage[utf8]{inputenc}
\usepackage{textcomp}
\usepackage{eurosym}
\usepackage{graphicx}
\usepackage{multirow}
\usepackage{epstopdf}
\usepackage{epsfig}
\usepackage{stmaryrd}
\usepackage{multirow,url}
\usepackage{color,cite}
\usepackage{slashbox}
\usepackage{amsthm}
\usepackage{bm}

\usepackage{multirow}
\usepackage{caption}

\begin{document}

\title{Bringing SDN to the Mobile Edge}

\author{\IEEEauthorblockN{Konstantinos Poularakis$^\ddagger$, Qiaofeng Qin$^\ddagger$, Erich Nahum$^\diamond$, Miguel Rio$^{*}$, Leandros Tassiulas$^\ddagger$}
\\
\IEEEauthorblockA{$^\ddagger${Department of Electrical Engineering and Institute for Network Science, Yale University, USA}
\\
{$^{\diamond}${IBM T. J. Watson Research Center, USA}}
}
\\
{$^{*}${Department of Electrical and Electronic Engineering, University College London, UK}}
}

\maketitle

\thispagestyle{empty}

\begin{abstract}
Nowadays, Software Defined Network (SDN) architectures and applications are revolutionizing the way wired networks are built and operate. However, little is known about the potential of this disruptive technology in wireless mobile networks. In fact, SDN is based on a centralized network control principle, while existing mobile network protocols give emphasis on the distribution of network resources and their management. Therefore, it is challenging to apply SDN ideas in the context of mobile networks. In this paper, we propose methods to overcome these challenges and make SDN more suitable for the mobile environment. Our main idea is to combine centralized SDN and distributed control in a hybrid design that takes the best of the two paradigms; (i) global network view and control programmability of SDN and (ii) robustness of distributed protocols. We discuss the pros and cons of each method and highlight them in an SDN prototype implementation built using off-the-shelf mobile devices.
\end{abstract}

\section{Introduction} \label{section:introduction}

\subsection{Motivation}

Software Defined Network (SDN) technology promises to advance communication networks to a whole new level of programmability, which will allow to manage network resources on demand and at a granular level, and offer more flexible services to users~\cite{sdn-survey}. The main concept of SDN is to separate network control functions (control plane) from data forwarding devices (data plane) and shift them to a logically centralized and programmable network entity, the controller. Up to date, the vast majority of SDN studies refer to wired infrastructures, i.e., data center and ISP networks~\cite{sdn-b4},~\cite{poularakis-infocom17}. Little is known today about the potential of this disruptive technology in wireless mobile networks.

A major challenge that holds back the SDN penetration in mobile networks is their high level of volatility and the difficulty in establishing reliable communication between the control and data planes. For example, if the SDN controller loses connectivity to the mobile nodes, it will be impossible to reconfigure them, resulting in outdated (and clearly suboptimal) routing policies. Even if the controller is reachable, frequent network updates in the mobile network can take a long time and overload the controller.

The aforementioned problems are mainly because of the centralization of all the network control functions into the SDN controller. This centralized approach works well in wired networks which are relatively static and, therefore, the communication between the controller and the data plane nodes is much more stable and reliable than in the mobile counterpart. However, this approach can be problematic in the context of highly mobile networks.

At the same time, traditional (non-SDN) distributed routing protocols, such as AODV and OLSR~\cite{manet-survey}, have been shown to work well in mobile networks. These protocols can be used for network discovery and routing in presence of network failures and mobility, providing a robust network architecture. Their main limitation compared to SDN is their lack of global network view and programmability to realize end-to-end network policies. All the above mean that \emph{it would be beneficial to design a network architecture that combines the benefits of the two paradigms; (i) global network view and programmability of SDN and (ii) robustness of distributed routing protocols.}

\subsection{Methodology}

Motivated by the above discussion, we propose to revisit the strict separation between control and data planes of SDN.
Namely, we put forward a \emph{hybrid control design} in which network control is split between the SDN controller and the mobile devices (data plane).
That is, we allow the mobile devices to make their own data forwarding decisions in a distributed manner, without involving the SDN controller.
Some previous works have explored the issue of hybrid control to offload a central SDN controller, e.g., see \cite{difane}, \cite{devoflow}, \cite{lazyctrl}, but not in the context of mobile networks.

We need to emphasize that the co-existence of two different control planes in the same network (e.g., a centralized SDN and a distributed IP) poses risks for fault-free routing, such as forwarding loops and blackholes~\cite{routing-anomalies}. Therefore, we need to combine the two control planes in a way that ensures that the packets will reach their destinations in reasonable time and without faults. In this paper, we propose three alternative ways or methods to combine centralized SDN and distributed control planes.

{\bf{Method 1: Dynamic migration of control protocol}}.
In this method, a part of the nodes in the SDN network can dynamically migrate to a distributed IP protocol (e.g., AODV or OLSR) forgoing the SDN forwarding rules. For example, the nodes in a certain region in which frequent network changes happen can be selected to act this way. By its nature, the distributed protocol will adapt to these changes faster than the remote SDN controller would do. However, deciding \emph{which specific part of the network to migrate and when}, as well as \emph{how the nodes that run a distributed protocol will interact with their neighbors that run an SDN protocol} are non-trivial open problems.

{\bf{Method 2: Cluster-based hierarchical control}}.
In the second method, we propose to partition the mobile network into several clusters and run a distributed IP protocol for each cluster independently from the rest.
The distributed protocols will manage the traffic routing within the clusters (level 1 of the hierarchy), while the centralized SDN controller will be responsible for the traffic routing in the overlay network formed by the clusters (level 2 of the hierarchy). This way, many of the network changes will be handled by the distributed protocols, without involving the SDN controller. The latter will only need to \emph{coordinate the distributed protocols} when the routing path spans more than one cluster. An important challenge here is to design the clustering, i.e., \emph{how many and how big the clusters should be}.

{\bf{Method 3: Distribution of backup SDN rules}}.
In this method, we \emph{proactively distribute ``backup'' SDN forwarding rules} to the mobile nodes to specify how they should alter their forwarding behavior if network conditions change. For example, we may store a backup rule to a node advising it to forward packets to a different next-hop node if a certain link fails. The mobile node can use its local control logic to decide to follow the backup rule over the primary rules depending on its observation of the network conditions. It may also communicate with other nodes to synchronize its view of the network and coordinate its decisions with them. A challenge in realizing this scheme is that the storage, bandwidth and computing resources of the mobile nodes are typically limited, and therefore it may not be possible to support all possible network failure scenarios.

All the above methods require us to relax the main SDN concept of completely ``dumb'' data plane nodes which only follow the controller instructions. Instead, we need to push to mobile nodes some level of control logic. With this power at hand, the mobile nodes can decide to run distributed protocols independently (method 1) or in coordination (method 2) with the central SDN controller. They may even pre-store and dynamically activate backup forwarding rules so as to deviate from controller instructions in response to rapid network changes (method 3).

To highlight the benefits of the proposed approach, we implement an SDN prototype of a hybrid control architecture and execute experiments measuring the performance and limitations.
Our prototype implementation consists of common smartphone and laptop devices which are set up with OpenVSwitch (OvS) OpenFlow datapath~\cite{ovs}.
We find that by pushing a certain level of control logic to the mobile devices we can provide a multi-fold reduction in failure reaction time compared to the ``pure'' (fully-centralized) OpenFlow system where the controller responds to all failures. The gains highly depend on the quality of the wireless channel between the controller and the mobile devices.

The contributions of this work are summarized as follows:
\begin{itemize}

\item \emph{Hybrid SDN Control}. We revisit the separation between the control and data planes of SDN architecture to make it more suitable for the mobile network environment. By pushing a level of control logic to the mobile nodes, we make the SDN architecture more robust and adaptive to network changes.

\item \emph{Centralized and distributed control combination}. We propose three specific methods to combine centralized SDN and distributed control planes. We describe the interaction between the two planes and discuss the key challenges in realizing them.

\item \emph{Proof-of-Concept Prototype Implementation.} We implement a hybrid SDN prototype and execute experiments measuring the performance and limitations. We find that the hybrid SDN system can provide a multi-fold reduction in failure reaction time compared to the ``pure'' (fully-centralized) OpenFlow system for a range of experiments. The gains depend on the quality of the wireless channel between the controller and the mobile devices.

\end{itemize}

The rest of the paper is organized as follows. After presenting some background on SDN in Section \ref{section:background}, we describe the proposed hybrid SDN control methods in Section \ref{section:methods}. Section \ref{section:implementation} presents the prototype implementation and experimental results. We conclude our work in Section \ref{section:conclusion}. 
\thispagestyle{empty}

\section{Background} \label{section:background}

In this section, we present a brief background on SDN in wireless mobile networks. We then discuss some existing works targeting to improve the robustness of the SDN architecture.

\subsection{SDN in wireless mobile networks}

The majority of SDN works target wired networks such as ISP networks and data centers~\cite{sdn-survey}. Few recent works have attempted to apply these ideas to wireless networks. For example, \cite{softcell} proposed to exploit SDN and shift control functions from core gateways to middleboxes. This can eliminate management bottlenecks by decentralizing the network operation. On the other hand, \cite{softran} suggested the deployment of software defined radio access networks (RAN). The idea is to assign the management of multiple base stations to a global controller and with this unified control improve performance. An interesting point is the suggestion for splitting the control decisions to those requiring full information (hence assigned to the global controller) and to those that need fast response (assigned to the radio elements). By allowing the dynamic split of the above decisions depending on the network conditions, adaptive and flexible control in the RAN can be achieved \cite{flexran}. Going a step further to the data plane, \cite{openradio} proposed to turn the base stations to fully programmable nodes facilitating this way the virtualization of network resources.

The above developments manifest that small, yet solid steps have been made towards designing software defined wireless networks. Interestingly, recently there have been efforts to deploy SDN soft switches even to handheld mobile devices, see \cite{android-ovs,meSDN,syrivelis}. By doing so, programmatic control of the traffic between the mobile devices can be achieved. These works manifest the actual potential for deploying SDN-enabled mobile networks.

\subsection{Robustness of SDN architecture}

Conventional SDN protocols such as OpenFlow fully rely on the controller to reconfigure the data plane nodes. This fully centralized approach can be problematic in mobile networks due to their volatile nature and the lack of reliable communication between the controller and data plane nodes. A rather straightforward method to increase controller availability is to deploy multiple (logically centralized but physically distributed) controllers throughout the network~\cite{robust_cp}. However, this approach incurs high deployment cost and cannot cover all cases. In contrast, we believe that the robustness problem is inherent to the centralized nature of SDN, and hence the solution lies on revisiting the strict separation between the control and data planes. Some previous works have explored the issue of hybrid control to offload a central SDN controller, e.g., see \cite{difane}, \cite{devoflow}, \cite{lazyctrl}, but not in the context of mobile networks.

\section{Combining centralized \& distributed control} \label{section:methods}
\thispagestyle{empty}

In this section, we propose three methods that can be used to combine centralized and distributed control to make SDN architecture more robust, and hence more suitable to the mobile network context.

\subsection{Migration to a distributed protocol}

\begin{figure}[t]
	\begin{center}
		\includegraphics[scale=0.6]{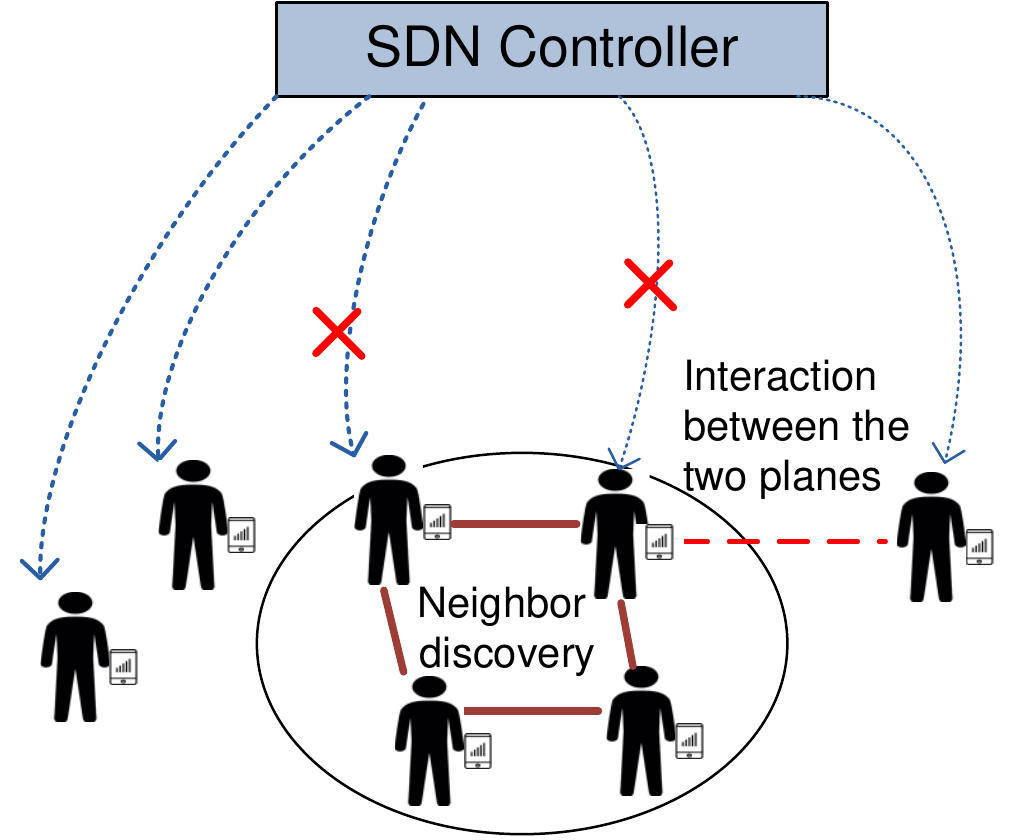}
		\caption{Example of migration from SDN to distributed control.}
		\label{fig:migration}
	\end{center}
\end{figure}

A natural method to increase robustness in a mobile SDN network is to use a distributed routing protocol ``\emph{as a backup}''.
That is, instead of relying on the remote controller to update network policies at all times, the mobile nodes can dynamically migrate from SDN to a distributed routing protocol (e.g., OLSR). Consider for example the mobile network in Figure \ref{fig:migration}. The nodes in the the circle may temporarily lose connectivity to the SDN controller due to  mobility and poor channel quality. To maintain connectivity to their neighbors and establish routing paths, these nodes can run a neighbor discovery process in a distributed manner. When conditions change, e.g., the connectivity to the controller becomes more stable, the nodes can be configured again by the controller and migrate back to SDN.

A drawback of this approach is that neighbor discovery and routing path computation processes typically take time to execute and hence they may result in slow network updates.
A way to expedite network updates is to make mobile nodes execute the distributed protocols long before the migration takes place.
Therefore, upon the migration time, the list of neighbor nodes and the respective routing paths will be already known and available to use by the distributed protocol.
Although this approach would reduce delay, it would also induce significant overheads since it would flood the network with discovery messages.
Hence, there is a \emph{tradeoff between delay and bandwidth consumption which can be tuned by the way distributed control execution is scheduled}.

Another challenge in this context is to ensure the proper interaction between the nodes that run the SDN and distributed control protocols.
It is well known in the literature that the co-existence of two different control planes in the same network (e.g., a centralized SDN and a distributed IP) poses risks for fault-free routing, such as forwarding loops~\cite{routing-anomalies}. For example, two neighbor nodes that run different protocols could ``ping-pong'' the same packet forever.
The SDN controller needs to identify and heal such routing anomalies, by appropriately configuring the SDN nodes that are adjacent to the nodes running the distributed protocol.

\subsection{Cluster-based hierarchical control}

Another method which has been shown to be quite effective in improving scalability and robustness of distributed routing protocols is \emph{clustering}~\cite{manet-cluster1}, \cite{manet-cluster2}. The main idea here is to allow nodes with close proximity to work together as a cluster and determine routes independently from other nodes.
The same concept has been successfully applied to ISP networks with the OSPF and BGP protocols~\cite{bgp}. Indeed, as these networks get big, they start using hierarchy to manage size and complexity. While part of that split has to do with administrative ownership, it also has to do with size (e.g., an ISP can have multiple autonomous systems).

\begin{figure}[t]
	\begin{center}
		\includegraphics[scale=0.38]{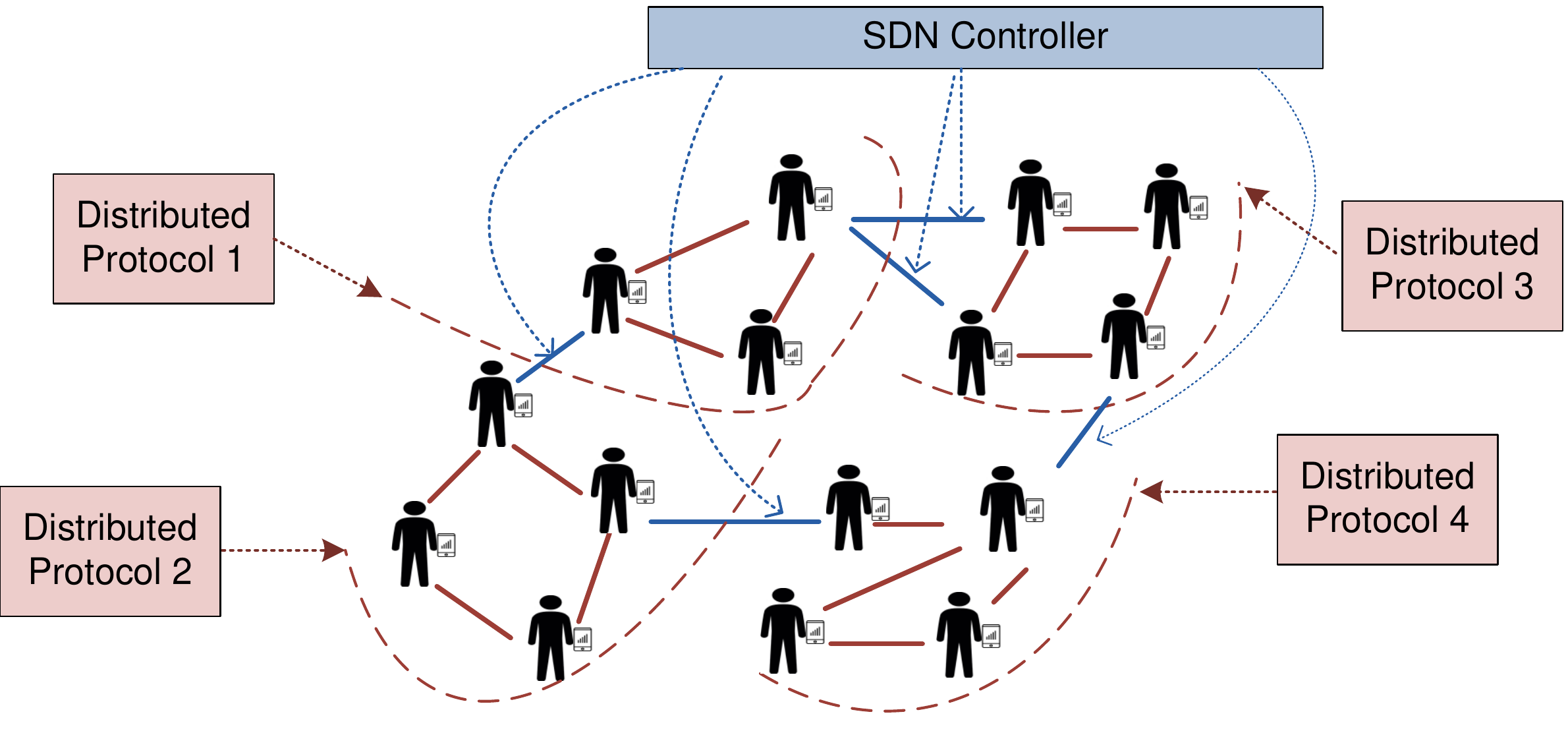}
		\caption{Example of clustering for hybrid SDN control.}
		\label{fig:cluster}
	\end{center}
\end{figure}

Inspired by these schemes, we propose to partition the mobile network into multiple clusters as it is depicted in Figure \ref{fig:cluster}.
Here, each cluster runs its own distributed protocol which is responsible to route traffic inside the cluster, or to the border nodes of neighbor clusters.
The role of the SDN controller is to ``guide'' the distributed protocols in routing traffic from one cluster to another.

To realize the above hybrid control scheme, the controller can compute and disseminate to the nodes the \emph{sequence of clusters} through which the packets must pass to reach their destination. This information can be encoded and stored in \emph{tags} attached to the packets. In fact, the OpenFlow protocol supports such kind of function by using VLAN tags or MPLS tags. Reading the tag's first entry helps the distributed protocol decide which neighbor cluster the packet should be sent to. After that, the tag is updated by removing the first entry, and so on, until the destination node is reached.

If there is only one cluster (the entire mobile network), then this corresponds to the traditional ``pure'' distributed protocol. In the other extreme case that each cluster consists of a single node, this corresponds to the ``pure'' SDN protocol. Intuitively, the larger the cluster sizes are the more network changes can be handled locally by the distributed protocols. However, the central SDN controller can only influence the traffic routing across multiple clusters. Hence, there is a \emph{tradeoff between controllability and robustness}.

The number and shape of clusters can be decided in a \emph{proactive} manner, i.e., ahead of the network changes. These decisions will have significant impact on the robustness of the SDN architecture. Periodically, the clustering solution can be refined to adapt to the new network environment (\emph{semi-proactive} scheme). Therefore, the critical questions are the following:
\begin{enumerate}
    \item In how many and how large clusters should the network be partitioned?

    \item How often and under what conditions should the clustering solution be updated?
\end{enumerate}

\subsection{Distribution of backup SDN rules}

A third method to increase robustness of SDN architecture is by storing \emph{``backup'' forwarding rules} to the mobile nodes. These rules will tell the nodes how they should change their forwarding behavior in response to network changes such as link or node failures. Having stored such rules, the mobile nodes can instantaneously and autonomously alter their behavior without asking the remote SDN controller. Similar schemes have been proposed in wired networks for handling a limited number of link failures~\cite{detour}, but the issue becomes more challenging in highly dynamic mobile networks.

\begin{figure}[t]
	\begin{center}
		\includegraphics[scale=0.48]{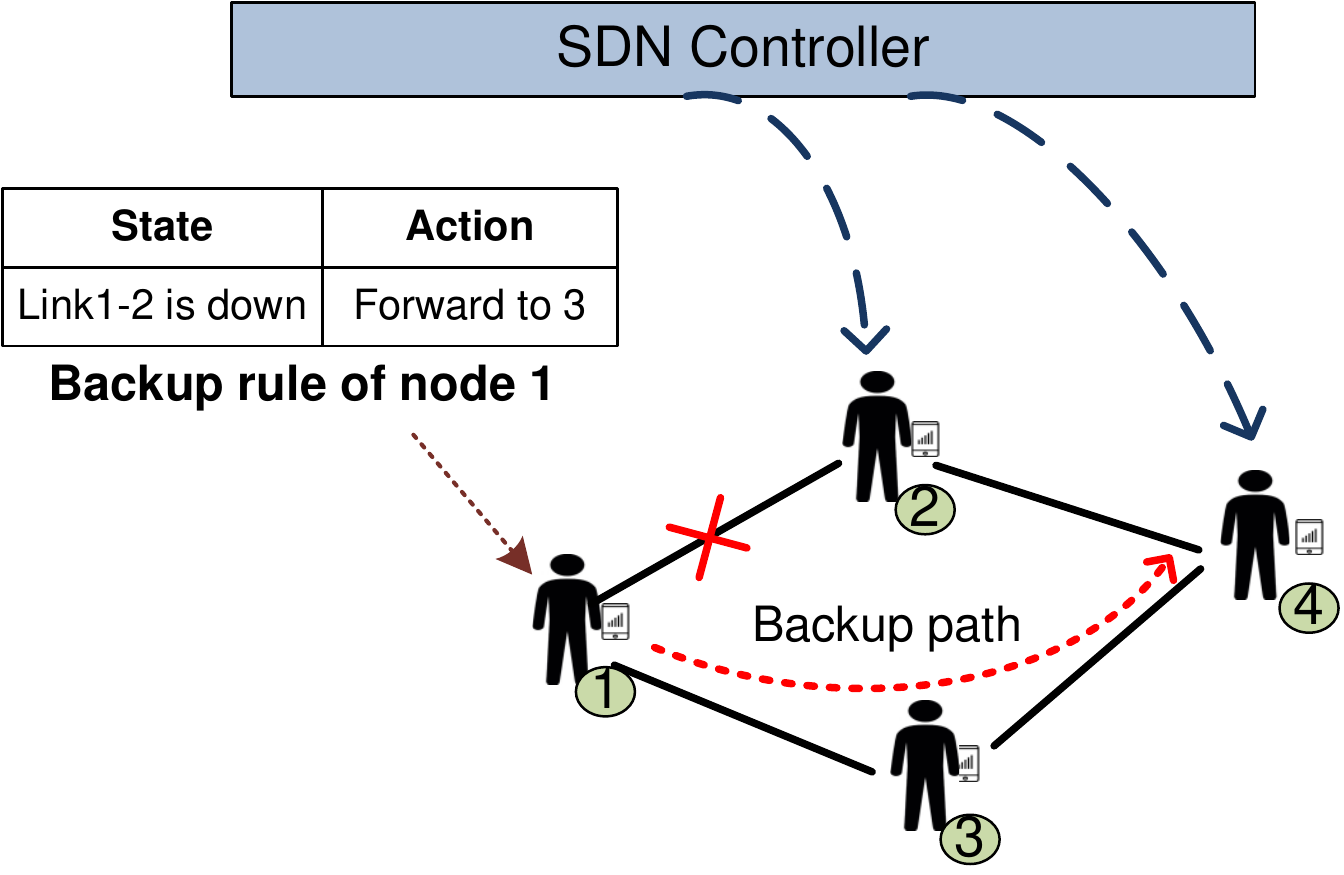}
		\caption{Example of backup rule storage.}
		\label{fig:backup}
	\end{center}
\end{figure}

Let us consider the example in Figure \ref{fig:backup}. The SDN controller pro-actively computes and stores a backup forwarding rule at node 1 advising how to overcome the failure of the link to node 2. In fact, recently developed extensions of OpenFlow support such functions. Namely, OpenState~\cite{openstate} allows data plane nodes to run state machines, which can use to keep information about the state of the network links and match packets against both their headers and the current state values. Hence, this method is inline with current technology standards.

We need to stress that the number of possible link failures and backup routing paths increase exponentially with the size of the network. Therefore, we may need to store a very large number of backup rules to support all the failures. At the same time, the memory size of the mobile nodes is limited and hence it may be insufficient to store all the rules.
On the positive side, there exist methods that compress the routing tables using wildcard rules and hence they can reduce the number of backup rules~\cite{compress}.

The mobile nodes can exchange messages to synchronize their local state information and make more efficient routing decisions. This way, for example, a node can learn that a link failure happened several hops away. However, such message exchange should be scheduled with caution to avoid flooding the network with messages and consuming the limited bandwidth resources of the mobile nodes. Therefore, the critical questions are the following:
\begin{enumerate}
    \item How many and which backup rules should be stored at each node to support which failures?

    \item How often and which nodes should exchange state information?
\end{enumerate}

\thispagestyle{empty}

\subsection{Discussion of the methods}
The three proposed methods combine SDN and distributed control paradigms in different ways. Method 1 explores the \emph{time dimension} and dynamically migrates some nodes from one protocol to another depending on the network conditions. Method 2 explores the \emph{space dimension} to split network control between the SDN controller and the distributed protocols based on the geographical position of the nodes. Finally, Method 3 explores the \emph{memory and processing resources of the mobile nodes} to proactively store and run backup forwarding rules without the controller involvement.

\begin{figure}[t]
	\begin{center}
		\includegraphics[scale=0.7]{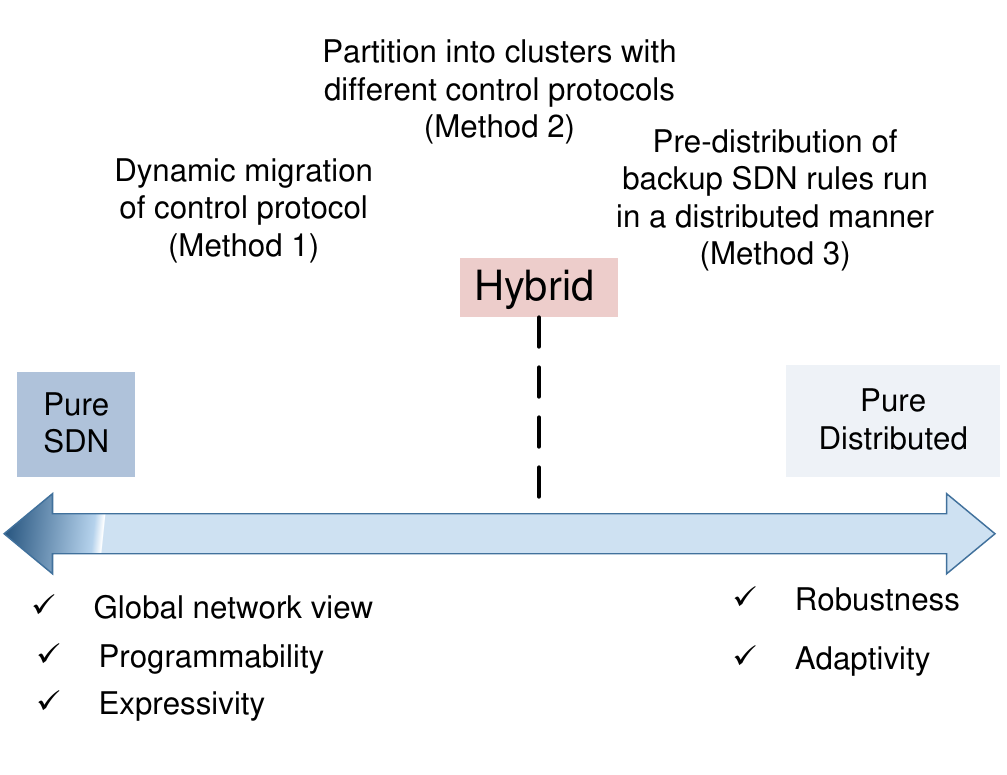}
		\caption{Operating point between SDN and distributed control.}
		\label{fig:operatingpoint}
	\end{center}
\end{figure}

The result of each method is a hybrid control protocol operating between pure SDN and pure distributed control paradigms (Figure \ref{fig:operatingpoint}).
An open research question is \emph{how to find the right operating point between the two paradigms to take the best of them}.
Intuitively, the more time is spent on running distributed protocols the closer to the pure distributed point Method 1 is. Similarly, Method 2 is close to the pure distributed point when few clusters of large sizes are formed. Finally, the more backup rules are stored at the mobile nodes, the more scenarios they can manage in a distributed manner ending up closer to  the pure distributed point.

\section{Proof-of-concept prototype implementation} \label{section:implementation}

In order to verify the feasibility of methods we proposed, we implement a prototype of a hybrid SDN system using mobile devices. The implementation consists of three parts. First, we design the architecture of a single mobile node, where both SDN and distributed control logics are enabled. Then, we set up a real mobile network with multi hops and alternative routing paths among the nodes. Last but not the least, we conduct measurements on the performance metrics that are related to the hybrid control method.

\thispagestyle{empty}

\subsection{Node Architecture}
In order to be close to realistic cases, our testbed consists of modern off-the-shelf mobile devices. Namely, we use the Nexus 4 smartphone (quad-core 1.5 GHz processor, 2 GB of RAM) with Android 4.2.1 system. The first necessary part we install is \emph{Open vSwitch (OvS)}. With OvS, the smartphone is turned into a virtual switch and can thus receive control messages and forwarding rules from SDN controller. Another significant part is a \emph{local software agent}. In all the three methods we propose, the mobile node should be able to modify the forwarding table locally in order to take distributed actions. Therefore, we design a local agent which can communicate with the virtual switch in OpenFlow protocol. What is more, the local agent should also have other functions. More specifically, in the distributed protocol migration method and the cluster-based control method, the local agent should maintain a distributed protocol, which involves synchronizing network states with other nodes, as well as calculating routing paths. In the third method, the local agent should store a set of backup forwarding rules and decide when to enable them. To verify such functions, we program a simple application of local agent for experimentation. It periodically sends "heartbeat" messages to neighbor nodes to judge whether the link is up or down. Once a link failure is detected, it will forward packets involved to its another interface. Such an application can be a basic operation in all of the three proposed methods.

\begin{figure}[t]
	\begin{center}
		\includegraphics[scale=0.28]{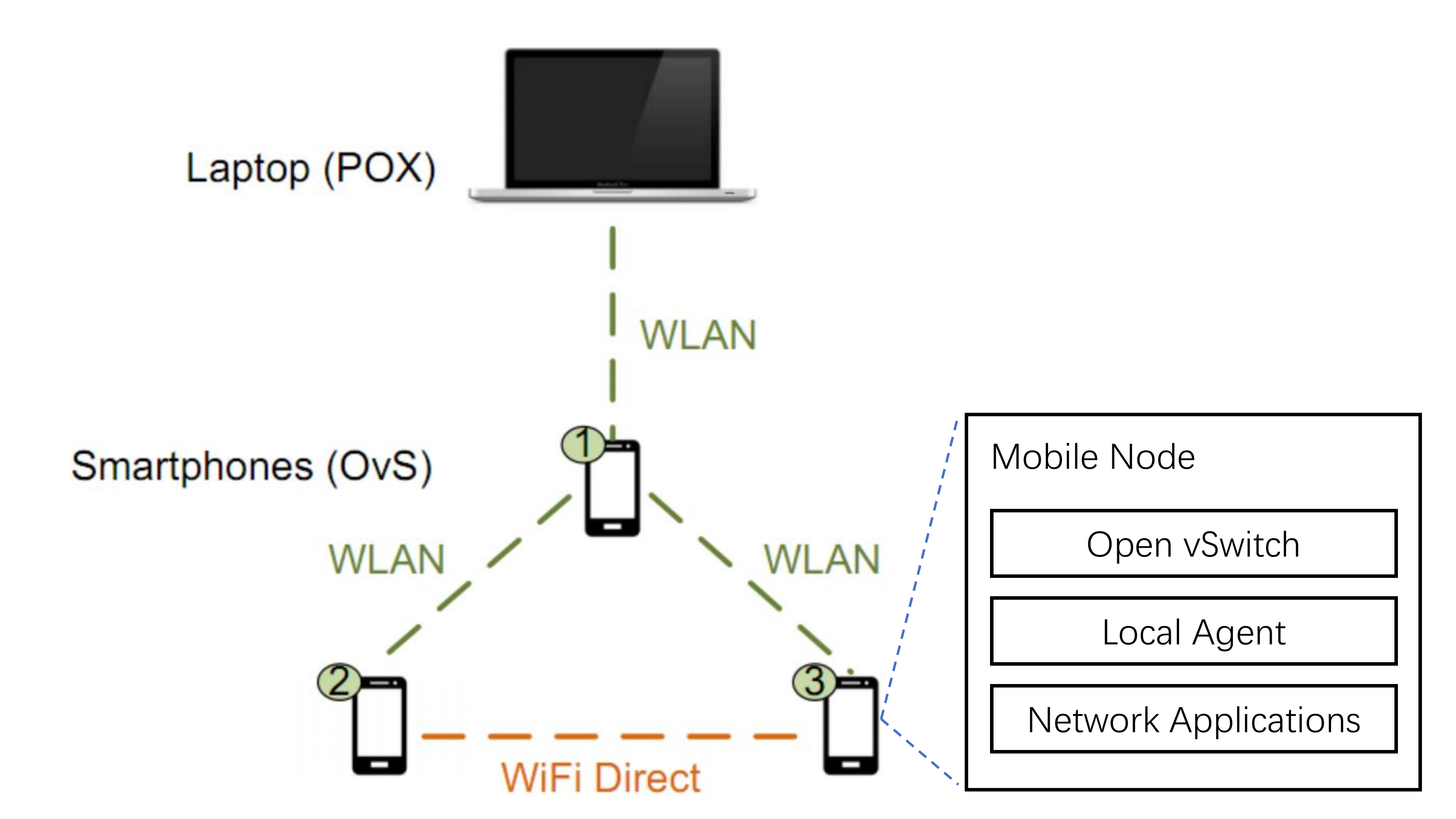}
		\caption{The network topology involving three smartphones and the SDN controller installed on a laptop.}
		\label{fig:topo}
	\end{center}
\end{figure}

\subsection{Network Setup}
We deploy smartphones to form a mobile network. The network should contain multiple paths between a pair of nodes, so that it becomes important to make routing decisions. In most modern Android smartphones including Nexus 4, the wireless ad hoc mode is no longer supported. However, due to the multiple network interfaces a smartphone has (e.g. 3G, Wi-Fi, Bluetooth), it is still common to have such networks. In the scenario we concern, the network consists of three Nexus 4 phones and one Macbook laptop. The laptop works as the central SDN controller. We use POX \cite{pox} to build our controller instance. For the network connections, first, one smartphone works as a Wi-Fi access point (hotspot), and other smartphones as well as the laptop connect to it. At the same time, the last two smartphones enable WiFi direct protocol between them and therefore have a second routing path as well as a different set of IP and MAC addresses. The network topology is shown in Figure \ref{fig:topo}. In a case such as one link is down, the flow between the two nodes can be migrated to another path by conducting network address translation under the control of either the SDN controller or the local agent.

\subsection{Measurements on Delay}
\begin{figure}[t]
	\centering
        \includegraphics[width =6.9cm, height=4.6cm]{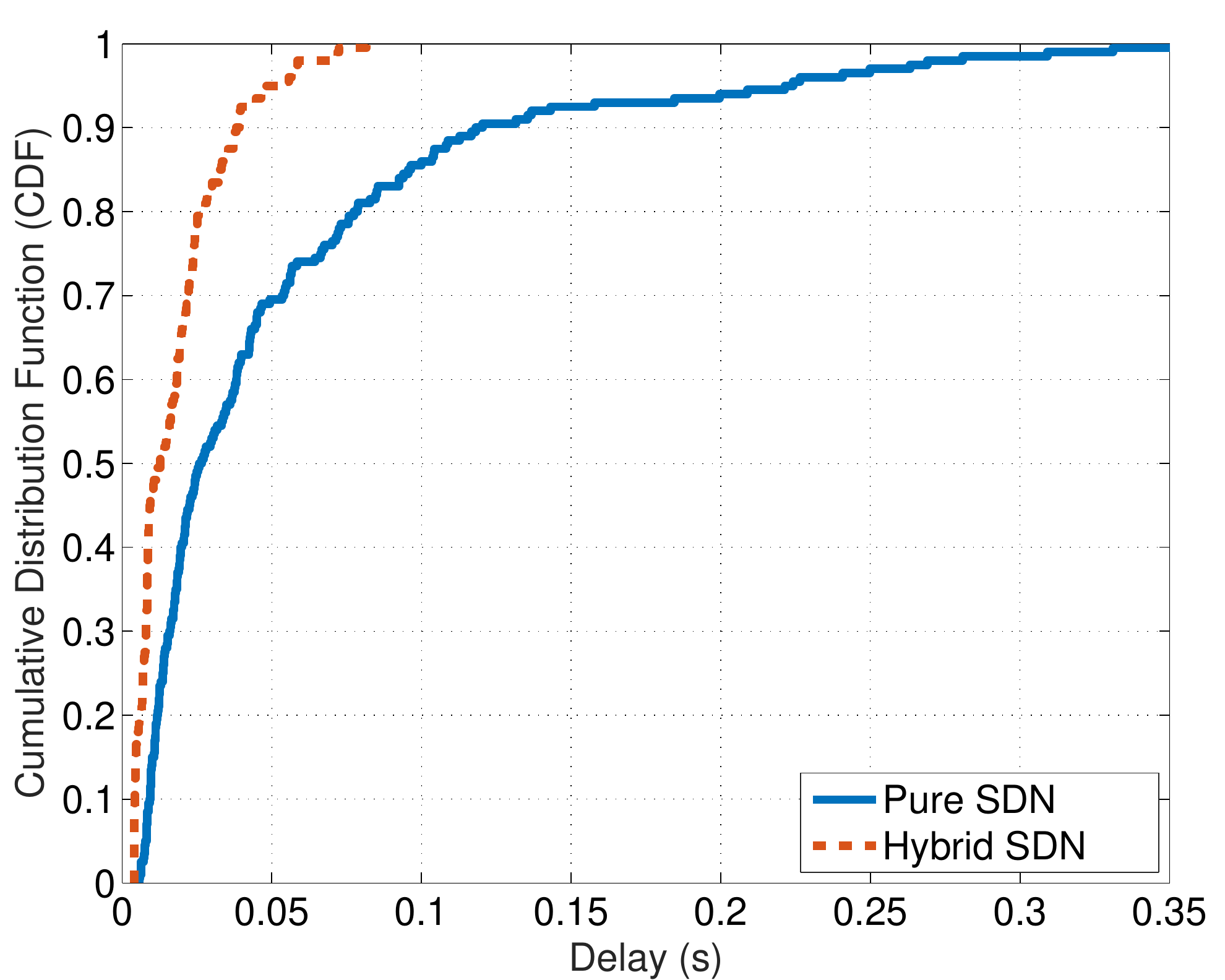} \label{fig:measurementa}
	\caption{CDF of delay values of pure and hybrid SDN control.}
	\label{fig:implement}
\end{figure}

Compared with wired networks, a mobile network has high mobility that leads to a rapidly changing topology. The SDN controller should react to these changes quickly. However, links in mobile networks are generally less stable, making delay a more significant factor. In methods we proposed, the local agent should totally or partly take over the network control from the central controller at the proper time. Therefore, it is important to compare the delay of the central control, which is limited by the link capacity between the controller and the mobile node, and the local control, whose bottleneck is the limited calculating capacity of the mobile node (smartphone). We simulate a link failure between smartphone 2 and 3, and new forwarding rules are sent to OvS by the SDN controller (in pure SDN solution) or the local agent (in hybrid SDN solution). The delay is defined as the time interval from when the failure is detected to when new rules are received and set up by OvS. We measure this delay for 200 times, and have a CDF plot of measured delay. Figure \ref{fig:implement} shows that both the average value and the variance of delay are smaller in the hybrid SDN case. These results strongly support the feasibility and advantages of the combination of centralized control and distributed control.

\thispagestyle{empty}

\section{Conclusion} \label{section:conclusion}
In this paper we proposed methods to make SDN more suitable to the mobile environment.
Our key observation is that many of the robustness problems of SDN are inherent to its fully centralized nature.
Therefore, we can benefit by revisiting the strict separation between the control and data planes and pushing some control logic to the mobile nodes in a hybrid design.
As a proof-of-concept, we programmed off-the-shelf mobile devices so as to be capable of totally or partly taking over the network control from the central controller at the proper time.

An open research question is how to find the right operating point between SDN and distributed control for each of the proposed methods.
This requires to solve challenging optimization problems, such as clustering and forwarding rule storage, as well as to design new protocols for the interaction between the two control paradigms. Besides, extensive testbed-based evaluations are necessary, similar to those presented, in order to identify all possible trade-offs and performance limitations of such systems.

\section*{Acknowledgement}
This research was sponsored by the U.S. Army Research Laboratory and the U.K. Ministry of Defence under Agreement Number W911NF-16-3-0001. The views and conclusions contained in this document are those of the authors and should not be interpreted as representing the official policies, either expressed or implied, of the U.S. Army Research Laboratory, the U.S. Government, the U.K. Ministry of Defence or the U.K. Government. The U.S. and U.K. Governments are authorized to reproduce and distribute reprints for Government purposes notwithstanding any copy-right notation hereon.

\end{document}